\documentclass{IEEE_lsens}
\usepackage{textcomp}
\usepackage{graphicx}
\usepackage{subfigure}
\usepackage{array}
\newcolumntype{C}[1]{>{\centering\arraybackslash}p{#1}}
\usepackage{epsfig}
\usepackage{epstopdf}
\usepackage{amsmath,graphicx}
\usepackage{tikz}
\usepackage{cite}
\usepackage{multicol}
\usepackage{cleveref}

\ifCLASSINFOpdf
\else
\fi
\usepackage[T1]{fontenc} 
\usepackage{amsmath}
\usepackage[cmintegrals]{newtxmath}
\usepackage{bm}

\usepackage{comment}
\usepackage{array}
\usepackage{url}
\ifCLASSINFOpdf
\else
\fi
\providecommand{\hypersetup}[1]{\relax}

\hypersetup{pdftitle={ImAiR},
pdfsubject={Typesetting},
pdfauthor={Ayush Tripathi},
pdfkeywords={Class, IEEE, IEEE\_lsens, IEEE Sensors Letters, LaTeX, Typesetting, TeX}}

\begin{document}

\IEEELSENSarticlesubject{Sensor signal processing}

\title{ImAiR: Airwriting Recognition framework using Image Representation of IMU Signals}

\author{\IEEEauthorblockN{Ayush Tripathi\IEEEauthorrefmark{1}, Arnab Kumar Mondal\IEEEauthorrefmark{2}, 
Lalan Kumar\IEEEauthorrefmark{1,3}\IEEEauthorieeemembermark{1}, Prathosh A.P.\IEEEauthorrefmark{4}}
\IEEEauthorblockA{\IEEEauthorrefmark{1}Department of Electrical Engineering,  Indian Institute of Technology Delhi, New Delhi, India\\
\IEEEauthorrefmark{2}Amar Nath and Shashi Khosla School of Information Technology,  Indian Institute of Technology Delhi, New Delhi, India\\
\IEEEauthorrefmark{3}Bharti School of Telecommunication,  Indian Institute of Technology Delhi, New Delhi, India\\
\IEEEauthorrefmark{4}Department of Electrical Communication Engineering, Indian Institute of Science, Bengaluru, India\\
\IEEEauthorieeemembermark{1}Member, IEEE}}

\IEEEtitleabstractindextext{%
\begin{abstract}
The problem of Airwriting Recognition is focused on identifying letters written by movement of finger in free space. It is a type of gesture recognition where the dictionary corresponds to letters in a specific language. In particular, airwriting recognition using sensor data from wrist-worn devices can be used as a medium of user input for applications in Human-Computer Interaction (HCI). Recognition of in-air trajectories using such wrist-worn devices is limited in literature and forms the basis of the current work. In this paper, we propose an airwriting recognition framework by first encoding the time-series data obtained from a wearable Inertial Measurement Unit (IMU) on the wrist as images and then utilizing deep learning-based models for identifying the written alphabets. The signals recorded from 3-axis accelerometer and gyroscope in IMU are encoded as images using different techniques such as Self Similarity Matrix (SSM), Gramian Angular Field (GAF) and Markov Transition Field (MTF) to form two sets of 3-channel images. These are then fed to two separate classification models and letter prediction is made based on an average of the class conditional probabilities obtained from the two models. Several standard model architectures for image classification such as variants of ResNet, DenseNet, VGGNet, AlexNet and GoogleNet have been utilized. Experiments performed on two publicly available datasets demonstrate the efficacy of the proposed strategy. The code for our implementation will be made available at https://github.com/ayushayt/ImAiR.
\end{abstract}

}

\maketitle

\section{Introduction}

\subsection{Background}

The process of writing letters in open space with unrestrained finger movements is known as airwriting\cite{7322243}. It can be utilized to provide a user with a rapid and touch-free input alternative that can be used in Human-Computer Interaction applications\cite{4154947}. Over the last few years, the challenge of recognizing writing from motion sensors has received significant attention, and a variety of techniques have been developed for the task\cite{amma2014airwriting, 8272739, patil2016handwriting, alam2020trajectory}. The studies employing motion sensors can be separated into two categories: dedicated devices, such as a wearable glove\cite{10.1145/2540048}, Wii remote\cite{li2021cross,xu2021novel}, smartphone\cite{li2018deep}, and vision-based studies\cite{roy2018cnn}. However, these solutions require the user to carry an additional physical device, which may be inconvenient for the users. To counteract this, the second category of techniques make use of wearable gadgets such as a ring worn on the index finger\cite{jing2017wearable} and smartbands\cite{yanay2020air,SCLAiR,abir2021deep} to recognize airwriting. 

\subsection{Related Work}

Several attempts have been made to recognize palm movements using wrist-worn devices\cite{10.1145/3264929, wen2016serendipity}. However, most of the research in this field has focused on the scenario in which a stable and flat surface is employed during the writing process, resulting in hand stabilization. This results in reduced noise because of presence of visual and haptic feedback to the user. In this work, airwriting recognition using a wrist-worn device is addressed. However, it is to be noted that majority of the movement during writing process is contributed from the user's palm, whereas the wearable is present on the wrist. This makes the task of airwriting recognition from wrist-worn devices challenging. The problem of airwriting recognition using a wrist-worn device was first addressed in \cite{7444820}. However, the work considered only a single subject with Dynamic Time Warping (DTW) as distance measure for classification, thus making it to be user dependent. A Convolutional Neural Network (CNN) based user-independent framework was proposed in \cite{yanay2020air}, for detecting airwritten English uppercase alphabets. In \cite{abir2021deep}, the authors explored several techniques for interpolating signals to make them of fix-length in addition to a 2D-CNN based classification scheme. All these studies make use of single stage classification with cross-entropy loss minimization. A $2$-stage classification scheme based on Supervised Contrastive Loss was suggested in \cite{SCLAiR} which yielded state-of-the-art airwriting recognition performance on an in-house and a publicly available airwriting dataset. All the studies performed hitherto have been focused on directly using time-series signals. All the studies related to airwriting recognition using motion sensor data have been focused on using the time series directly for classification using different Deep Learning models. Due to the scarcity of pre-trained architectures for time-series, there haven't been any attempts in exploring transfer learning techniques for the task.
By encoding the time-series to image representations, the powerful 2D-CNN models pre-trained on the ImageNet dataset can be utilized. These enable learning the complex attributes present in the encoded image representation and form the core idea of the current work.

\begin{figure*}[!ht]
  \centering
  \centerline{\includegraphics[width=0.85\linewidth]{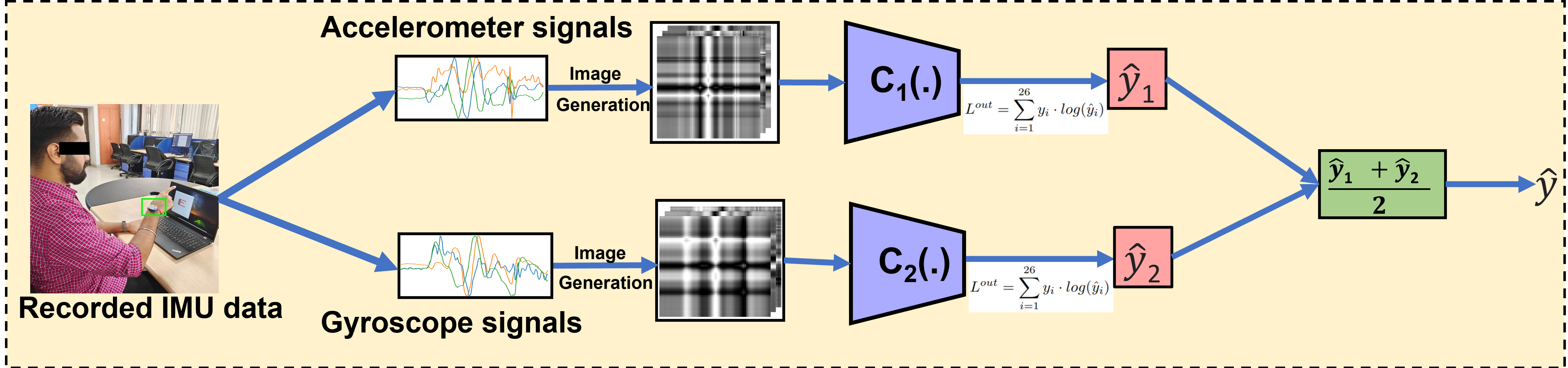}}
  \caption{Block diagram depicting the proposed method. } 
\label{fig:block_diagram}
\end{figure*}

\subsection{Objectives and Contributions}

In this work, an airwriting recognition framework using image representation of time-series data is proposed for the first time. The time-series data is collected through 3-axis accelerometer and gyroscope sensors from an Inertial Measurement Unit (IMU). Given that the airwriting recognition task requires the identification of fine-grained movements of the wrist to be identified, the sensitivity of IMU plays a huge role in this task. The required sensitivity is readily available in off-the-shelf smart bands and can therefore be used for HCI applications. By encoding the time-series to image representation, the capabilities of CNN were used to extract more diversified characteristics from complicated local patterns in the images. Given this intuition, the proposed framework consist of encoding time series data from accelerometer and gyroscope into 2 separate set of images. Then, a 2D-CNN based classifier is trained by using individual feature vectors, separately. The posterior probabilities from the two classifiers for each class are averaged out and the class with maximum probability is chosen as the predicted alphabet. We validate the efficacy of the proposed approach by performing experiments on 2 different publicly available datasets \cite{yanay2020air,SCLAiR}. For the purpose of image encoding, several techniques – Self Similarity Matrix, Gramian Angular Field, and Markov Transition Field have been explored and their suitability for airwriting recognition has been evaluated. Various standard architectures from the image domain - ResNet\cite{resnet}, DenseNet\cite{densenet}, VGGNet\cite{vggnet}, GoogleNet\cite{googlenet}, and AlexNet\cite{alexnet}, and their variants have been explored for the classifier. 

\section{Proposed Method}

The ImAiR framework proposed in this paper starts with the acquisition of data using IMU sensors during the process of airwriting as shown in Figure \ref{fig:block_diagram}. The signals from 3-axis accelerometer and gyroscope are then separated and encoded into images using different image encoding techniques to form two sets of 3-channel images corresponding to the accelerometer and gyroscope respectively. These encoded images are then fed into two separate CNN based classification models for classifying the alphabet. Finally, the posterior probabilities corresponding to the different classes from both the classifiers are averaged and alphabet prediction is made based on this average probability score.  

\subsection{Problem Description}

A multivariate time-series recorded using 3-axis accelerometer and gyroscope of an Inertial Measurement Unit (IMU) placed on the wrist of the dominant hand of a human subject while writing uppercase English alphabets forms the input for the method. The input can be represented as a $L \times 6$ matrix $X$, where $L$ is the number of time steps and $6$ is the number of sensor axes. Each individual time-series $x_j (j \in \{1,...,6\})$ is then encoded into image representation by using the function $f(.) : x_j \rightarrow \mathcal{I}_j$. This results in a $L \times L \times 6$ dimensional representation $\mathcal{I}$ of the input data $X$. The image representation can be divided into two parts - $\mathcal{I}_A$ and $\mathcal{I}_G$ corresponding to the accelerometer and gyroscope respectively each of the two having a dimension of $L \times L \times 3$. The classifier network, $C_1(.)$ maps the input $\mathcal{I}_A$ to a $26$ dimensional vector representing the probabilistic distribution $P(\hat{y}_A=c_i|\mathcal{I}_A;\Theta_A)$. Here, $\hat{y}_A$ is the predicted class from classifier $C_1$, $c_i = \{A,B,....,Z\}$ is the set of classes corresponding to each of the $26$ alphabets, and $\Theta_A$ is the set of parameters for the classifier. Similarly, classifier network, $C_2(.)$ maps the input from the gyroscope $\mathcal{I}_G$ to a probabilistic distribution $P(\hat{y}_G=c_i|\mathcal{I}_G;\Theta_G)$. Subsequently, we compute the average class probabilities as:

\begin{equation}
    P(\hat{y}=c_i|\mathcal{I}_A,\mathcal{I}_G) = \frac{P(\hat{y}_A=c_i|\mathcal{I}_A;\Theta_A) + P(\hat{y}_G=c_i|\mathcal{I}_G;\Theta_G)}{2}
\end{equation}

Based on the computed average probability, the input $X$ is assigned to class $\hat{y} = \arg \max_{c_i}  P(\hat{y}=c_i|\mathcal{I}_A,\mathcal{I}_G)$.

\begin{table*}[!t]
\caption{Recognition accuracies using different Image Encoding techniques and model architecture combinations on Dataset-1. Entries in blue denote an accuracy greater than 85\% and bold entries denote the best achieved accuracy for each image encoding technique.}
\centering
\scalebox{0.7}{
\begin{tabular}{|l|lll|lll|lll|lll|}
\hline 
 & & \textbf{SSM} &  & & \textbf{MTF} & & & \textbf{GASF} &    & & \textbf{GADF} &                             \\\hline\hline
                      \textbf{Architecture}& \textbf{Accelerometer} & \textbf{Gyroscope} & \textbf{Fusion} & \textbf{Accelerometer} & \textbf{Gyroscope} & \textbf{Fusion} & \textbf{Accelerometer} & \textbf{Gyroscope} & \textbf{Fusion} & \textbf{Accelerometer} & \textbf{Gyroscope} & \textbf{Fusion} \\\hline
ResNet18              & 0.7648                 & 0.7933             & 0.8241          & 0.7051                 & 0.7308             & 0.7711          & 0.7485                 & 0.7315             & 0.7923          & 0.8099                 & \textcolor{blue}{0.8537}             & \textcolor{blue}{0.8730}          \\
ResNet34              & 0.8065                 & 0.8032             & 0.8419          & 0.6937                 & 0.7417             & 0.7754          & 0.7477                 & 0.7626             & 0.8138          & 0.7995                 & \textcolor{blue}{0.8561}             & \textcolor{blue}{0.8663}          \\
ResNet50              & 0.7850                 & 0.8118             & 0.8342          & 0.6942                 & 0.7328             & 0.7713          & 0.7393                 & 0.7521             & 0.8027          & 0.8152                 & \textcolor{blue}{0.8609}             & \textcolor{blue}{0.8779}          \\
ResNet101             & 0.8133                 & 0.7995             & 0.8439          & 0.7014                 & 0.7552             & 0.7832          & 0.7641                 & 0.7309             & 0.8128          & 0.8115                 & \textcolor{blue}{0.8653}             & \textcolor{blue}{0.8776}          \\
ResNet152             & 0.8118                 & 0.7974             & 0.8484          & 0.6954                 & 0.7685             & 0.7894          & 0.7622                 & 0.7480             & 0.8089          & 0.8010                 & \textcolor{blue}{0.8547}             & \textcolor{blue}{0.8756}          \\
VGG11                 & 0.7421                 & 0.7650             & 0.8080          & 0.6458                 & 0.7026             & 0.7395          & 0.6969                 & 0.7070             & 0.7689          & 0.7769                 & 0.8308             & 0.8456          \\
VGG11-BN             & 0.7776                 & 0.7887             & 0.8299          & 0.6757                 & 0.7144             & 0.7576          & 0.7150                 & 0.7320             & 0.7856          & 0.7971                 & 0.8335             & \textcolor{blue}{0.8569}          \\
VGG13                 & 0.7578                 & 0.7756             & 0.8226          & 0.6619                 & 0.7205             & 0.7516          & 0.7145                 & 0.7070             & 0.7689          & 0.7824                 & 0.8263             & 0.8496          \\
VGG13-BN             & 0.7651                 & 0.8005             & 0.8330          & 0.6802                 & 0.7468             & 0.7778          & 0.7190                 & 0.7296             & 0.7884          & 0.7779                 & 0.8385             & 0.8542          \\
VGG16             & 0.7704                 & 0.7952             & 0.8326          & 0.6968                 & 0.7248             & 0.7703          & 0.7209                 & 0.7270             & 0.7868          & 0.7976                 & 0.8362             & \textcolor{blue}{0.8597}          \\
VGG16-BN             & 0.8089                 & 0.8048             & 0.8489          & 0.7397                 & 0.7692             & \textbf{0.8046}          & 0.7526                 & 0.7549             & 0.8056          & 0.8106                 & 0.8439             & \textcolor{blue}{0.8653}          \\
VGG19                 & 0.7858                 & 0.7809             & 0.8315          & 0.7099                 & 0.7556             & 0.7959          & 0.7332                 & 0.7197             & 0.7858          & 0.8034                 & 0.8304             & \textcolor{blue}{0.8586}          \\
VGG19-BN              & 0.7834                 & 0.8080             & 0.8429          & 0.7142                 & 0.7617             & 0.7971          & 0.7535                 & 0.7484             & 0.8094          & 0.8212                 & 0.8289             & \textcolor{blue}{0.8650}          \\
DenseNet121           & 0.8120                 & 0.7937             & 0.8441          & 0.7036                 & 0.7244             & 0.7723          & 0.7443                 & 0.7651             & 0.8092          & 0.8250                 & 0.8451             & \textcolor{blue}{0.8711}          \\
DenseNet161           & 0.8238                 & 0.8099             & \textbf{0.8526}          & 0.7462                 & 0.7438             & 0.7903          & 0.7785                 & 0.7638             & \textbf{0.8215}          & 0.8379                 & \textcolor{blue}{0.8621}             & \textcolor{blue}{\textbf{0.8843}}          \\
DenseNet169           & 0.7795                 & 0.8287             & 0.8455          & 0.7149                 & 0.7362             & 0.7778          & 0.7412                 & 0.7465             & 0.8067          & 0.8294                 & \textcolor{blue}{0.8617}             & \textcolor{blue}{0.8778}          \\
DenseNet201           & 0.8048                 & 0.8039             & 0.8479          & 0.6915                 & 0.7701             & 0.7892          & 0.7773                 & 0.7569             & 0.8138          & 0.8316                 & 0.8414             & \textcolor{blue}{0.8725}          \\
AlexNet               & 0.6757                 & 0.7041             & 0.7521          & 0.5038                 & 0.5494             & 0.6044          & 0.5581                 & 0.6154             & 0.6631          & 0.7176                 & 0.7834             & 0.8072          \\
GoogleNet             & 0.7791                 & 0.7877             & 0.8296          & 0.6472                 & 0.6629             & 0.7149          & 0.7342                 & 0.7055             & 0.7783          & 0.8034                 & \textcolor{blue}{0.8559}             & \textcolor{blue}{0.8600}      \\ \hline 
\end{tabular}
}
\label{tab:Dataset1results}
\end{table*}

\subsection{Image Encoding Methodologies}
There can be various choices of the time-series to image encoding techniques and classifier architectures that can be adopted without any constraints.

\subsubsection{Self Similarity Matrix}

Self-Similarity Matrix (SSM) is a graphical representation of similar sequences in a time-series. For a time-series $V = \{v_1,v_2,....,v_L\}$ of length $L$, the entries of SSM are computed as:

\begin{equation}
    SSM(i,j) = f(v_i,v_j) \;\;\; ; i,j \in \{1,2,....,L\}
\end{equation}

Here, $f(v_i,v_j)$ is a measure of similarity the two points $v_i$ and $v_j$. In our work, we use Euclidean distance as the similarity measure.

\subsubsection{Gramian Angular Field}
Granular Angular Field (GAF) \cite{wang2015imaging} is a method to encode time series to an image by using polar coordinate based matrix, while still preserving the inherent temporal relationships in the time-series. For $V = \{v_1,v_2,....,v_L\}$, a time-series of length $L$, the first step is to rescale the elements of $V$ so that they fall within the range of $[-1,1]$. The rescaled time-series is further converted to polar coordinates as $\theta_i = arccos(\tilde{v_i})$ and $r_i = \frac{t_i}{L}$.

Here, $t_i$ is the timestamp and the division by $L$ is done in order to regularize the span in the polar coordinate system. Rescaling the time-series ensures that there is a proper inverse mapping from $(r,\theta)$ to $(\tilde{v},t)$ system. Subsequently, temporal correlation between samples at different timestamps can be identified by using either trigonometric sum or difference of angles ($\theta$). Gramian
Angular Summation Field (GASF) is defined by using the cosine of sum of angles as:

\begin{equation}
    GASF = 
    \begin{bmatrix} 
    cos(\theta_1 + \theta_1) & \dots & cos(\theta_1 + \theta_L) \\
    cos(\theta_2 + \theta_1) & \dots & cos(\theta_2 + \theta_L) \\
    \vdots & \ddots & \vdots\\
    cos(\theta_L + \theta_1) & \dots & cos(\theta_L + \theta_L) \\
    \end{bmatrix}
\end{equation}

Similarly, Gramian Angular Difference Field (GADF) can defined by using the sine of difference of angles as:

\begin{equation}
    GADF = 
    \begin{bmatrix} 
    sin(\theta_1 - \theta_1) & \dots & sin(\theta_1 - \theta_L) \\
    sin(\theta_2 - \theta_1) & \dots & sin(\theta_2 - \theta_L) \\
    \vdots & \ddots & \vdots \\
    sin(\theta_L - \theta_1) & \dots & sin(\theta_L - \theta_L) \\
    \end{bmatrix}
\end{equation}

\subsubsection{Markov Transition Field}

Markov Transition Field (MTF) \cite{wang2015imaging} is another technique to encode the dynamic transition statistics of a time series by representing the Markov transition probabilities in a sequential manner. The first step in generating MTF is to sample elements of a time-series $V = \{v_1,v_2,....,v_L\}$ into $Q$ bins. Further, each sample $v_i$ is assigned to its corresponding bin $q_j$ ($j \in [1,Q]$). Using this mapping, a $Q\times Q$ dimensional adjacency matrix $W$ is constructed, the elements $w_{ij}$ of which are given by the frequency with which a point in bin $q_j$ is followed by a point in bin $q_i$. The elements of $W$ are then used to construct the MTF matrix as:  

\begin{equation}
    MTF = 
    \begin{bmatrix} 
    w_{ij|v_1 \in q_i,v_1 \in q_j} & \dots & w_{ij|v_1 \in q_i,v_L \in q_j} \\
    w_{ij|v_2 \in q_i,v_1 \in q_j} & \dots & w_{ij|v_2 \in q_i,v_L \in q_j} \\
    \vdots & \ddots & \vdots \\
    w_{ij|v_L \in q_i,v_1 \in q_j} & \dots & w_{ij|v_L \in q_i,v_L \in q_j} \\
    \end{bmatrix}
\end{equation}

\subsection{Model Architecture}

We explored various different standard architectures for the classifier block: ResNet\cite{resnet} ($18$, $34$,$50$,$101$, and $152$ layer variants), DenseNet\cite{densenet}($121$, $161$, $169$, and $201$ variants), VGGNet\cite{vggnet}(with the number of weight layers equal to $11$, $13$, $16$, and $19$; with and without Batch Normalization) , GoogleNet\cite{googlenet}, and AlexNet\cite{alexnet}. The parameters of the model were initialized by using pre-trained weights from the ImageNet classification task and are optimized by minimizing the Cross-Entropy Loss using the Adam optimizer.

\section{Experiments and Results}

\subsection{Datasets}

\begin{itemize}
    \item \textbf{Dataset-1} - We used the publicly available dataset from \cite{yanay2020air} that consists of recordings obtained from $55$ subjects while writing English uppercase letters ($15$ repetitions). Data was recorded using a Microsoft band 2 at the maximum sampling rate supported by the device of $62$ Hz.
    
    \item \textbf{Dataset-2} - We used another publicly available dataset \cite{SCLAiR} which comprised of recordings obtained from 20 subjects during $10$ repetitions of writing English uppercase alphabets. Signals were recorded using a Noraxon Ultium EMG sensor (with an internal IMU) \cite{noraxon} placed on the wrist. The sampling rate was $200$ Hz and the signals were downsampled to $62$ Hz. Performance details of the used IMU are presented in Table \ref{tab:IMUperformance}.
\end{itemize}

\begin{table}[!h]
\caption{Performance details of the used IMU in Dataset-2.}
\centering
\scalebox{1}{
\begin{tabular}{ccc}
\hline
\textbf{}              & \textbf{Range}  & \textbf{Sensitivity}     \\ \hline
\textbf{Accelerometer} & $\pm$16g             & 2048 LSB/g               \\
\textbf{Gyroscope}     & $\pm$2000 degree/sec & 16.4 LSB/(degree/second) \\ \hline
\end{tabular}
}

\label{tab:IMUperformance}
\end{table}

\subsection{Experimental Details}

Similar to previous studies \cite{yanay2020air,SCLAiR}, we applied a two-step preprocessing strategy to the recorded IMU signals. First, the length of each signal was made fixed (to $155$) by discarding samples if the signal was of greater length or padding zeros otherwise. Subsequently, Z-normalization was applied to each of the $6$ signals individually.

In the first set of experiments, we evaluated the performance of combination of different image encoding techniques and classification models for the task of airwriting recognition on Dataset-1. For this purpose, we split the data from $55$ subjects into two parts - training data comprising of recordings from $40$ subjects (40*26*15 = 15600 samples) and test set from remaining $15$ subjects(15*26*15 = 5850 samples). Such a split ensures that the evaluation is performed in a user independent manner. The training data was further divided into a training set and a validation set having 80:20 ratio. A mini-batch training process was employed while keeping the batch size equal to $32$ and early stopping with a patience of $10$ epochs. 
In the next set of experiments, we performed leave-one-subject-out (LOSO) validation on both Dataset-1 and Dataset-2. We used the DenseNet161 model architecture for classification using all four image encoding techniques: SSM, MTF, GASF, and GADF. The performance of the proposed ImAiR framework in this setting is compared with state-of-the-art performance on both the datasets.

\subsection{Results and Comparison}

\begin{figure}[!t]
    \centering
    \centerline{\includegraphics[width=0.75\linewidth]{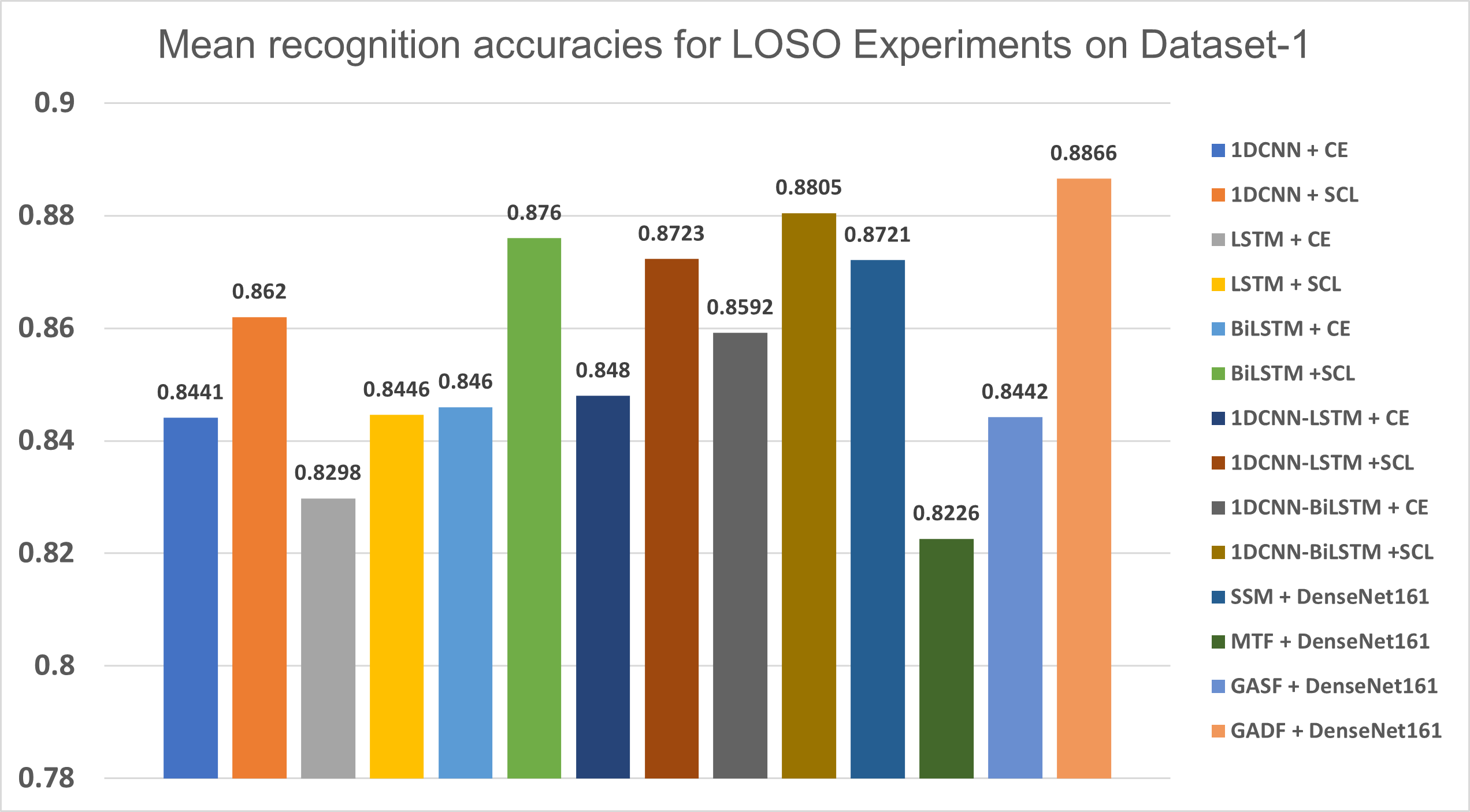}}
    \caption{Recognition accuracies for LOSO experiment on Dataset-1}
    \label{fig:LOSO_dataset1}
\end{figure}

\begin{figure}[!t]
    \centering
    \centerline{\includegraphics[width=0.75\linewidth]{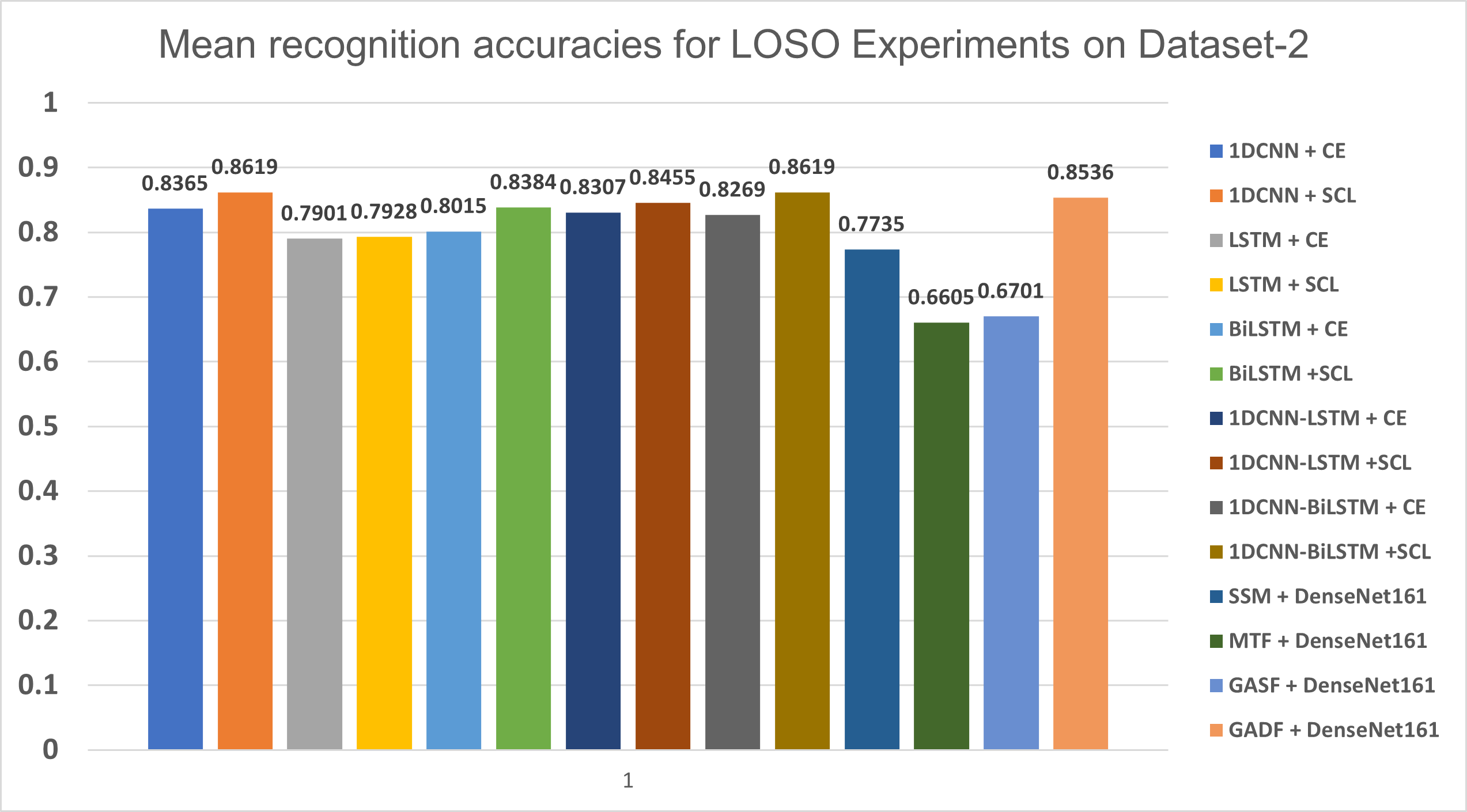}}
    \caption{Recognition accuracies for LOSO experiment on Dataset-2}
    \label{fig:LOSO_dataset2}
\end{figure}

Table \ref{tab:Dataset1results} lists the performance of different model architectures and image encoding techniques for the first set of experiments on Dataset-1. The results of the two individual classifiers (using argmax over $P(\hat{y}_A=c_i|\mathcal{I}_A;\Theta_A)$, $P(\hat{y}_G=c_i|\mathcal{I}_G;\Theta_G)$) and their average ($P(\hat{y}=c_i|\mathcal{I}_A,\mathcal{I}_G)$) for accelerometer and gyroscope based classification using different image encoding techniques are presented. It may be noted that DenseNet161 classifier provides the best recognition accuracy in all the scenarios. Among the different encoding methods, the performance of GADF is superior for all different classifiers. The best performance is obtained on using the GADF+DenseNet161 combination and the achieved accuracy is $88.43\%$. 

In Figures \ref{fig:LOSO_dataset1} and \ref{fig:LOSO_dataset2}, the results for LOSO experiments performed on Dataset-1 and Dataset-2 are presented and compared with state-of-the-art techniques that include different Deep Learning architectures (1DCNN, LSTM, BiLSTM, 1DCNN-LSTM, and 1DCNN-BiLSTM) with Cross Entropy and Supervised Contrastive Loss based techniques. Performing the experiments in a LOSO setting allows us to directly compare our results with those of \cite{yanay2020air} and \cite{SCLAiR}. It is observed that on Dataset-1, by using the GADF+DenseNet161 combination, we achieve an accuracy of $88.66\%$, which is the best reported accuracy on the given dataset. On Dataset-2, the best accuracy achieved is $85.36\%$, which is comparable to state-of-the-art. The dip in mean recognition accuracy on Dataset-2 in comparison to Dataset-1 may be attributed to the small amount of samples (4940 samples) available for training the model which may not be sufficient to learn the complex attributes present in the encoded image representation.

\section{Conclusion}
In this paper, we explored an airwriting recognition framework by using encoded image representation of time-series data from accelerometer and gyroscope sensors in a wrist-worn IMU. It is seen that the proposed approach outperforms the state-of the-art accuracy on a publicly available dataset \cite{yanay2020air}, while also performing significantly well on a another dataset \cite{SCLAiR}. Future work may be focused on exploring techniques that are well-suited to learn the attributes present in the signals from small datasets to increase the utility of the airwriting recognition framework for HCI applications.

\bibliographystyle{IEEEtran}
\bibliography{refs.bib}

\end{document}